# Highlights

## The influence of AGEs and enzymatic cross-links on the mechanical properties of collagen fibrils

Julia Kamml,Chun-Yu Ke,Claire Acevedo,David S. Kammer

- High contents of cross-links cause stiffening of the collagen fibril in the higher strain range.
- Sliding of tropocollagen molecules is inhibited when AGEs cross-links are dominating force transfer between tropocollagen molecules.
- Load transfer to tropocollagen molecules is causing stiffening of the fibril.
- Reduced sliding and stiffening of the collagen fibril lead to less energy dissipation in the tissue
- Impaired tissue properties can be caused by strengthening and stiffening at the collagen fibril level

# The influence of AGEs and enzymatic cross-links on the mechanical properties of collagen fibrils

Julia Kamml[a], Chun-Yu Ke[b], Claire Acevedo[c,d] and David S. Kammer[a,*]

[a]*Institute for Building Materials, ETH Zurich, Switzerland*
[b]*Department of Engineering Science and Mechanics, Pennsylvania State University, University Park, PA, USA*
[c]*Department of Mechanical Engineering, University of Utah, Salt Lake City, Utah, USA*
[d]*Department of Biomedical Engineering, University of Utah, Salt Lake City, Utah, USA*

ABSTRACT

Collagen, one of the main building blocks for various tissues, derives its mechanical properties directly from its structure of cross-linked tropocollagen molecules. The cross-links are considered to be a key component of collagen fibrils as they can change the fibrillar behavior in various ways. For instance, AGEs (Advanced-Glycation Endproducts), one particular type of cross-links, have been shown to accumulate and impair the mechanical properties of collagenous tissues, whereas enzymatic cross-links (ECLs) are known for stabilizing the structure of the fibril and improving material properties. However, the reasons for whether a given type of cross-link improves or impairs the material properties remain unknown, and the exact relationship between the cross-link properties and fibrillar behavior is still not well understood. Here, we use coarse-grained steered molecular models to evaluate the effect of AGEs and ECLs cross-links content on the deformation and failure properties of collagen fibrils. Our simulations show that the collagen fibrils stiffen at high strain levels when the AGEs content exceeds a critical value. In addition, the strength of the fibril increases with AGEs accumulation. By analyzing the forces within the different types of cross-links (AGEs and ECLs) as well as their failure, we demonstrate that a change of deformation mechanism is at the origin of these observations. A high AGEs content reinforces force transfer through AGEs cross-links rather than through friction between sliding tropocollagen molecules, which leads to failure by fracture of the tropocollagen molecules. We show that this failure mechanism, which is associated with lower energy dissipation, results in more abrupt failure of the collagen fibril. Our results provide a direct and causal link between increased AGEs content, inhibited intra-fibrillar sliding, increased stiffness, and abrupt fibril fracture. Therefore, they explain the mechanical origin of bone brittleness as commonly observed in elderly and diabetic populations. Our findings contribute to a better understanding of the mechanisms underlying impaired tissue behaviour due to elevated AGEs content and could enable targeted measures regarding the reduction of specific collagen cross-linking levels.

## 1. Introduction

Aging and diabetes impair mechanical properties and healing capacities in human tissues (Moseley, 2012; Couppé et al., 2016; Fox et al., 2011; Grant et al., 1997). However, precise mechanisms explaining such behaviour at different tissue scales remain still unknown. Tissue mechanical properties are conferred by its complex hierarchical structure. The most important structural protein maintaining the mechanical function and structural integrity of most tissues is collagen (Fratzl, 2008). Amongst different forms arising from the collagen family, collagen type I is the most abundant in the human body, occurring in tendon, bone, skin, cornea, lung, and vasculature (Hulmes, 2008). Its main building components are polypeptide chains called tropocollagen (TC) molecules that self-assemble into staggered fibrils with a characteristic periodic banding pattern (see Fig. 1c). Post-translational enzyme-driven modifications stabilize the fibrillar structure via cross-linking between TC molecules with enzymatic cross-links (ECLs). With aging and other factors like diabetes, another type of cross-linking is formed via a glycation process, the so-called Advanced-Glycation-Endproducts (AGEs). It has been observed that an increased density of these AGEs cross-links alters the mechanical properties of the collagen fibrils on the nano-scale (Gautieri et al., 2017; Davies et al., 2019; Fessel et al., 2014). However, whether or how this increased cross-linking within the collagen fibril (intrafibrillar) and the precise nano-scale mechanisms are responsible for impaired tissue properties has not been revealed so far.

The two types of cross-links are known to present different characteristics. *ECLs* are located at the end of the TC molecules (Fig. 1d-e) and their quantity controls strength and stiffness in collagen (Svensson et al., 2013; Buehler, 2006a; Depalle et al., 2015). These cross-links are formed via an enzymatic reaction mediated by lysil oxidase mainly at the ends of the TC molecules (see Fig. 1d). Initially, immature divalent cross-links will be formed, which are then reacting to form mature trivalent cross-links (Bailey et al., 1998). While throughout the early life the concentration of immature and mature cross-links reduces and increases, respectively, the cumulative ECL content stabilizes after the age of 20 years at approximately 1.2 ECLs per TC molecule (Eyre et al., 1988; Saito et al., 1997). *AGEs*, the second type of cross-links, are non-enzymatic since they are formed during glycation of proteins in the helical regions between TCs. They are predominantly formed in hyperglycemic systems due to the increased presence of sugar (Paul and Bailey, 1996). All pathways of their formation reaction have not fully been revealed, as multiple reactions can be involved (Bailey, 2001), leading to different molecular structures. Several types of AGEs have been found as intermolecular cross-links (Avery and Bailey, 2006), linking the TC molecules within the collagen fibril. AGEs accumulate with age, due to the long half-life of collagen, *e.g.*, the number of pentosidine in bone was found to grow by a factor of 5 from age 10 up to 80 years (Saito et al., 1997). In tissue of diabetic individuals, the process of AGEs formation is accelerated due to augmented blood glucose levels. By far the most abundant AGEs cross-link is glucosepane, which reaches levels about 1000 times higher than the contents of other cross-link types. Specifically, one glucosepane cross-link in every five TC molecules was found in tissue of 90 years old patients, whereas the AGE content reaches levels of one cross-link every two molecules in diabetic individuals (Sell et al., 2005). Eight potential binding sites for glucosepane's interfibrillar cross-linking have been proposed by Gautieri et al. (2014), but it has not been possible to experimentally quantify the exact amount, type and distribution of all types of AGEs present in the fibril.

*Corresponding author
dkammer@ethz.ch (D.S. Kammer)
orcid(s): 0000-0003-3782-9368 (D.S. Kammer)

Despite limited knowledge about the type and quantity of AGEs in collagen fibrils, it is known that their presence is a major cause of dysfunction in mature collagenous tissues in elderly patients, which is accelerated in diabetic subjects because of higher levels of glucose leading to increased AGEs formation (Bailey et al., 1998; Schnider and Kohn, 1982). Collagen tissue with a longer half-life is specifically affected by AGEs formation and their increased occurrence correlates with inferior material properties, e.g., with an increased fracture risk in bone (Avery and Bailey, 2006; Karim and Bouxsein, 2016). Several experimental studies have investigated the relation between cross-link density and mechanical behaviour of collagenous tissue (Acevedo et al., 2018b; Yang et al., 2012; Svensson et al., 2013). Nevertheless, the underlying mechanisms of how cross-links and specifically AGEs influence fibrils at the nanoscale could not be revealed so far due to limitations of imaging techniques and the complexity of collagen.

In-silico modeling provides a tool to overcome these experimental limitations and presents an opportunity to reveal the biomechanical effects of enzymatic and non-enzymatic intrafibrillar cross-linking on the mechanical properties of collagen fibrils at the small scale and the tissue at the larger scale. For studying the structure and mechanical behaviour of TC molecules in collagen fibrils, full atomistic simulations could be used (Uzel and Buehler, 2011; Gautieri et al., 2014; Buehler, 2006a). For instance, *Buehler et al.* used 2D atomistic simulations to describe the mechanical properties of collagen depending on TC molecule length, amount of enzymatic cross-links and mineralization (Buehler, 2006b, 2008), but full-scale simulations are computationally far too expensive to study multi-molecule fibril mechanics in full three dimensions.

This is where coarse-grained modeling can overcome the limitations of computational cost by representing the mechanical properties of the TC molecules using data from full-atomistic simulation results.

In this paper, we build a 3D coarse-grained model of the collagen fibril using steered molecular dynamics, where the mechanical response of TC molecules and cross-links are derived from reactive molecular dynamics simulations with atomistic resolution. With our simulations, we investigate the influence of cross-link density between TC molecules on stiffness, strength, and toughness (represented by work to failure) of collagen fibrils with a particular focus on AGEs increase on top of normal enzymatic cross-links. We show that higher densities of AGEs cause fibrilar stiffening (when fibrilar strain reaches a critical value of $\varepsilon_0 \approx 0.15$) and affect the failure mechanisms by causing the breaking of the TC molecules which results in abrupt failure of the fibril.

## 2. Material and methods

We build a full-scale model of a collagen fibril including enzymatic and AGEs cross-links at various densities. Both types of cross-links are considered because AGEs mostly occur in aged or diabetic tissue where enzymatic cross-links are naturally present, since they have mostly been formed during growth and adult development. We simulate a destructive tensile test of collagen fibrils and investigate various quantities related to their deformation, failure, and energy dissipation and how these depend on various cross-link densities. Our model is a coarse-grained molecular model where TC molecules forming the collagen fibril are represented by particles arranged in a chain and interacting according to multi-body potentials (Buehler, 2006a,b; Depalle et al., 2015). This approach allows us to reach length scales of hundreds of nanometers and time scales of microseconds, which would not be possible in full-scale atomistic modeling due to high computational costs.

### 2.1. Geometry of the collagen fibril

Collagen fibrils are bundles of TC molecules that are aligned in the longitudinal direction in the collagen-specific five-staggered pattern with gap and overlap zones (see Fig. 1c). In-vivo, TC is formed of three polypeptide chains (see Fig. 1a) twisted into a right-handed triple-helical structure with a diameter of about $1.5\,nm$ and a length of $300\,nm$ (Bhattacharjee and Bansal, 2005). Three different domains are distinguished in the TC molecule: the central triple helical domain and the two non-helical telopeptide ends called C- and N-terminal (see Fig. 1a). The central domain is by far the largest, comprising of about 95 percent of the total length of the molecule. In fibrillar collagen, the TC molecules self-assemble into staggered fibrils with a characteristic repeating banding pattern. The periodicity of the pattern is $D = 67\,nm$, where TC molecules are staggered with respect to their neighbors by multiples of $D$ (see Fig. 1c). Collagen fibrils have a thickness of about 50 to a few hundred nanometers (Hulmes, 2008).

We build one representative region of a collagen fibril showing five gap and overlap zones as shown in Fig. 1e. First, the geometry of the single TC molecules was obtained from the Protein Data Bank entry 3HR2, the atomistic model of a collagen type I microfibril based on X-ray crystallography (Orgel et al., 2006). We represent several amino acids of one TC molecule by a single particle and the mechanical properties of the entire molecule by a chain of these particles (see Fig. 1a-b). The coordinates of the particles in the longitudinal direction are obtained by using spline-fitting along the TC molecule structure and then aligning the particles equidistantly along this spline. The distance between the particles is chosen as the equilibrium distance $r_0 = 14.0$ Å of the inter-particle potential, approximating the diameter of the TC molecule (Depalle et al., 2015). This results in 218 particles per TC molecule. The angles between the two outer particles of each particle triplet along the chain are serving as equilibrium angles $\phi_i$ in the coarse-grained model. After the cross-section of collagen fibrils is built by meshing a sphere with a diameter of $d = 20.2$ nm with a triangular mesh, 155 TC molecules are replicated and aligned along the longitudinal axis of the fibril, resulting in a cylindrical shape. In the longitudinal direction the geometry is implemented with five gap and overlap zones of the TC molecules, where the measure of the gap $0.6 \cdot D$ and the overlaps $0.4 \cdot D$ accordingly. Finally, the collagen fibril was extended with 40 additional particles at the end of each TC molecule in order to ensure smooth force transmission by constraining and strengthening the fibril where external forces are applied.

### 2.2. Insertion of enzymatic cross-links

Enzymatic cross-links occur as immature divalent cross-links or mature trivalent cross-links (see Fig. 1d-g) between the telopeptide ends of the TC molecules and the helical domain of adjacent molecules in collagen fibrils (Saito and Marumo, 2015). They are covalent bonds between lysine or hydroxylysine residues. We model divalent and trivalent cross-links by linking two and three different TC molecules, respectively (Depalle et al., 2015). We vary the content of enzymatic cross-links between 0, 25, 50, 75 and 100 percent, where 100 percent corresponds to two enzymatic cross-links per TC molecule (2 mol/mol), *i.e.* one at each end. For enzymatic cross-link contents of less than 100 percent, we distribute them randomly amongst all the telopeptide ends of the collagen fibril such that no telopeptide end has more than one cross-link. The molecule to link to is chosen to be the one closest to the respective telopeptide end. The process of enzymatic cross-link insertion is done after equilibration of the collagen fibril model, as discussed in more detail in Sec. 2.5.

### 2.3. Insertion of AGEs cross-links

The density of AGEs cross-links *in vivo* depends on the tissue type, where tissue with low turnover and higher age generally shows a higher amount of cross-links (Takahashi et al., 1995). Very little is known about the exact location of AGEs in collagen. AGEs are formed *in vivo* via non-enzymatic reactions. AGEs cross-links have several different molecular compositions and emerge between TC molecules in the helical regions, depending on the sugars and amino acids involved in their formation and their relative distance. In general, a sugar moiety is added between two proteins involving lysine-to-lysine or lysine-to-arginine residues of two TC molecules leading to covalent bonding (Paul and Bailey, 1996; Avery and Bailey, 2006). Due to its frequency of occurrence, we use glucosepane as a representative for AGEs cross-links in our simulations: it is the most prominent AGE, reaching levels of 2000 pmol/mol in collagen tissue of 90 years old patients and of ∼ 4500 pmol/mol in diabetic patients (Sell et al., 2005). Hence, there is one AGE every 5 molecules in aged collagen and one AGE every 2 molecules in diabetic collagen.

In our model, cross-links are inserted randomly between two neighboring TC molecules. Telopeptide ends of the TC molecules (consisting of four particles in our model) are excluded as potential binding sites. Random cross-link density is measured as the number of AGEs cross-links per TC molecule. The potential binding sites between particles of every TC molecule are extracted and then cross-links are randomly created. If the density is less than 1 cross-link per TC molecule, TC molecules where cross-links are created are randomly chosen and then cross-links are randomly created at potential binding sites. We use densities of 0, 0.50, 1, 2, 5, 10, and 40 AGEs cross-links per TC molecule, which goes beyond the above-mentioned measured values, since the exact numbers of different AGEs cross-links have not been quantified so far.

### 2.4. Parameterization of the force field for coarse-grained modeling

We now will summarize the various particle interactions applied in our coarse-grained model. We consider the total energy of the modeled system as follows

$$E_{total} = E_{bond} + E_{angle} + E_{inter}$$
$$= \sum_{bond} \Phi_{bond}(r) + \sum_{angle} \Phi_{angle}(\phi) + \sum_{inter} \Phi_{inter}(r) \quad (1)$$

where $E_{bond}$ is the bond energy due to stretching, $E_{angle}$ the three-body interactions energy due to bending, and $E_{inter}$ the pairwise interaction energy due to molecular interactions such as Van-der-Waals forces.

The force between two particles linked via a bond is given by

$$F_{bond}(r) = -\frac{\partial \Phi_{bond}(r)}{\partial r}, \quad (2)$$

where $\Phi_{bond}$ is the bond potential and $r$ is the radial distance. We model the nonlinear deformation behaviour of a single collagen molecule in tensile tests as reported previously (Buehler and Wong, 2006; Buehler and Gao, 2006) with a tri-linear relation (more details provided in Appendix 8.1),

The influence of AGEs and enzymatic cross-links on the mechanical properties of collagen fibrils

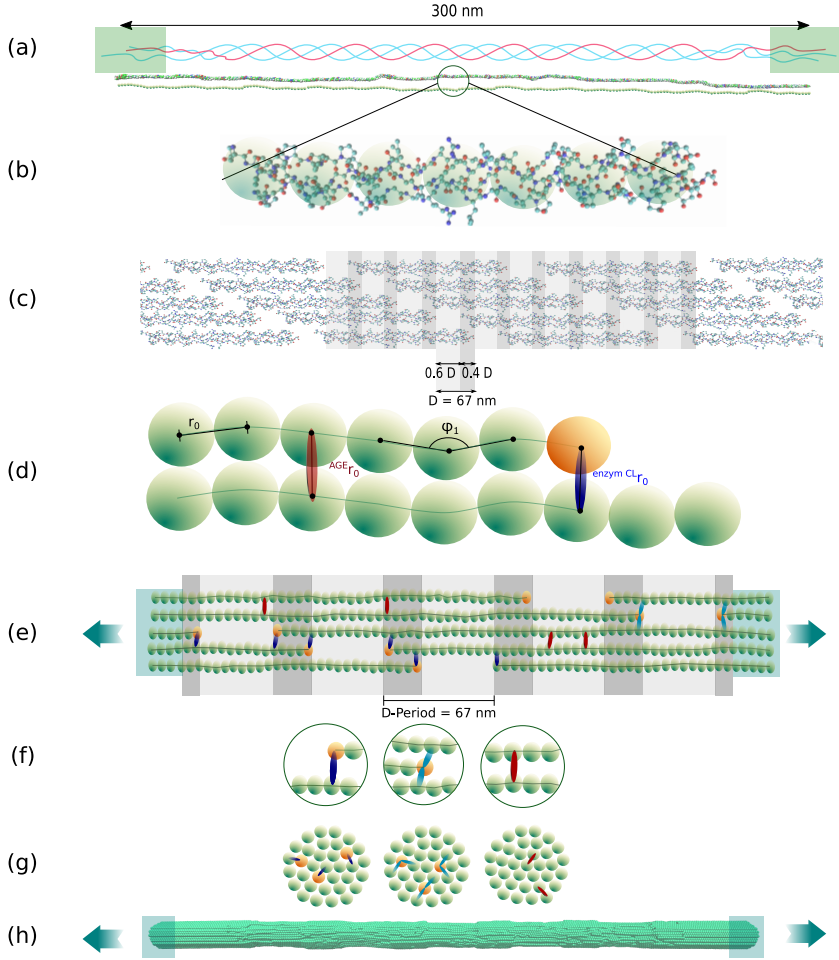

**Figure 1:** Schematic overview of creation of the coarse-grained model of collagen fibrils. (a) TC molecule with telopeptide ends and helical main region: consists of 3 $\alpha$-polypeptide helices twisted into each other. (b) Full-scale atomistic TC molecule is represented by a string of particles (coarse-grained model) - the forces between the particles represent the mechanical properties of the full-scale molecule. (c) Typical 5-staggering arrangement pattern of TC molecules within the collagen fibril with gap and overlap zones, leading the characteristic banding pattern. (d) The coarse-grained model parameters represent the mechanical properties of TC and the inter-molecular cross-links (AGEs and enzymatic cross-links). (e) Schematic representation of the configuration of the model as a representative part of the collagen fibril with 5 gap and overlap zones. Particles at the ends are rigidly constrained for applying force during displacement-controlled tensile tests using steered molecular dynamics. (f) Different types of bonds inserted in the model: Particles are linked via collagen covalent bonds within each TC molecule; ECLs as intermolecular covalent bonds (dark blue: divalent, light blue: trivalent) at the ends to the TC molecules, AGEs cross-links randomly placed between two TC molecules (red). (g) Cross sections with cross-links shown schematically. (h) Steered molecular dynamics tensile tests on collagen fibril model.

as follows

$$F_{bond}(r) = \begin{cases} -k_T^{(0)}(r - r_0) & \text{if } r < r_1 \\ -k_T^{(1)}(r - r_0) & \text{if } r_1 \leq r < r_{break} \\ z \cdot k_T^{(1)}(r - r_0) & \text{if } r_{break} \leq r < r_{break} + a \\ 0 & \text{if } r \geq r_{break} + a \end{cases} \quad (3)$$

where $r_0$ is the equilibrium distance between two particles linked by a bond, $k_t^{(0)}$ and $k_t^{(1)}$ are spring constants of the deformation, and $a$ is defined as $a = z \cdot (r_{break} - r_1)$. The chosen tri-linearity accounts for fracture of the bond, which occurs at $r_{break}$. The fracture is regularized with the $z$-factor to avoid any discontinuities and provide computational stability. The $z$-factor is chosen small enough, such that its finite value does not affect the simulation results.

Forces arising from angle bending between triplets of particles are defined as

$$F_{angle}(\phi) = -k_B(\phi - \phi_i) \cdot \phi, \quad (4)$$

with $\phi_i$ being the varying equilibrium angles obtained from the initial coarse grained geometry (see Sec. 2.1), and $k_B$ being the bending stiffness of the molecule (Depalle et al., 2015; Buehler and Wong, 2006; Buehler, 2006b).

The non-bonded interactions of particles are modeled with the Lennard-Jones potential with a soft core, which is

**Table 1**
Parameters used in coarse-grained molecular dynamics mesoscale model of collagen fibrils (Depalle et al., 2015)

| Components | Parameters | Value |
|---|---|---|
| Collagen molecules | Equilibrium particle distance ($r_0$, Å) | 14.00 |
| | Critical hyperelastic distance ($r_1$, Å) | 18.20 |
| | Bond breaking distance ($r_{break}$, Å) | 21.00 |
| | Tensile stiffness parameter ($k_0$, kcal mol$^{-1}$Å$^{-2}$) | 17.13 |
| | Tensile stiffness parameter ($k_1$, kcal mol$^{-1}$ Å$^{-2}$) | 97.66 |
| | Regularization factor ($z$, -) | 0.05 |
| | Equilibrium angle ($\phi_0$, degree) | 170.0 to 180.0 |
| | Bending stiffness parameter ($k_b$, kcal mol$^{-1}$rad$^{-2}$) | 14.98 |
| | Dispersive parameter ($\epsilon_{LJ}$, kcal mol$^{-1}$) | 6.87 |
| | Dispersive parameter ($\sigma_{LJ}$, Å) | 14.72 |
| | Soft core parameter ($\lambda$, -) | 0.9 |
| Particles at ends of TC molecules | *same parameters as collagen molecules except:* | |
| | Bond breaking distance ($r_{break}$, Å) | 70.00 |
| Divalent Cross-links | Equilibrium particle distance ($r_0$, Å) | 18.52 |
| | Critical hyperelastic distance ($r_1$, Å) | 20.52 |
| | Bond breaking distance ($r_{break}$, Å) | 23.20 |
| | Tensile stiffness parameter ($k_0$, kcal mol$^{-1}$Å$^{-2}$) | 0.20 |
| | Tensile stiffness parameter ($k_1$, kcal mol$^{-1}$Å$^{-2}$) | 41.84 |
| Trivalent Cross-links | Equilibrium particle distance ($r_0$, Å) | 18.52 |
| | Critical hyperelastic distance ($r_1$, Å) | 22.12 |
| | Bond breaking distance ($r_{break}$, Å) | 24.81 |
| | Tensile stiffness parameter ($k_0$, kcal, mol$^{-1}$ Å$^{-2}$) | 0.20 |
| | Tensile stiffness parameter ($k_1$, kcal, mol$^{-1}$ Å$^{-2}$) | 54.60 |
| AGEs Cross-links | Equilibrium particle distance ($r_0$, Å) | 18.52 |
| | Critical hyperelastic distance ($r_1$, Å) | 22.72 |
| | Bond breaking distance ($r_{break}$, Å) | 31.72 |
| | Tensile stiffness parameter ($k_0$, kcal, mol$^{-1}$Å$^{-2}$) | 0.1 |
| | Tensile stiffness parameter ($k_1$, kcal, mol$^{-1}$Å$^{-2}$) | 8.00 |
| | Mass of each mesoscale particle, atomic mass units | 1358.7 |

given by

$$F_{\text{non-bond}}(r) = \begin{cases} F_{LJ}(r) & \text{if } r \geq \lambda\sigma_{LJ} \\ F_{LJ}(\lambda\sigma_{LJ}) & \text{if } r < \lambda\sigma_{LJ} \end{cases} \quad (5)$$

where

$$F_{LJ}(r) = \frac{1}{r}\left[48\epsilon_{LJ}\left(\frac{\sigma_{LJ}}{r}\right)^{12} - 24\epsilon_{LJ}\left(\frac{\sigma_{LJ}}{r}\right)^{6}\right]. \quad (6)$$

In these equations, $\epsilon_{LJ}$ is the well depth between two particles, $\sigma_{LJ}$ is the distance at which the intermolecular potential between the two particles is zero and $\lambda$ is the parameter to adjust the critical force associated to the soft core.

The parameters for the coarse-grained model of the interactions in the TC molecules as well as for enzymatic cross-links are taken from full-scale simulations in literature and adjusted to our approach (Depalle et al., 2015; Buehler, 2006a). For the AGEs cross-links, we conducted steered molecular full-scale tensile test simulation and a literature study and extracted parameters for the observed mechanical behavior (details provided in Appendix 8.1). All interaction parameters used in the simulations are displayed in table 1.

### 2.5. Simulations

All coarse-grained molecular dynamics simulations on single collagen fibrils were performed in *LAMMPS* (Plimpton, 1995). Ten particles at the end of the system are constrained to a rigid body similar to a clamp (see Fig. 1e,h) and kept at a stable constant distance apart from each other. In this configuration and without any cross-links (yet), a first equilibration is performed, where, first, an energy minimization using steepest descent is performed followed by a conjugate gradient energy minimization. This sets the system to an energetically favorable configuration. Further, the system is equilibrated for 80 ns using an NVT ensemble at a temperature of 300 K with a time step of $\Delta t = 10$ fs. After equilibrium is reached, enzymatic and AGEs cross-links are inserted following the procedure described in Sec. 2.2 and 2.3. At this point, another equilibration is performed with the inserted cross-links for 10 ns.

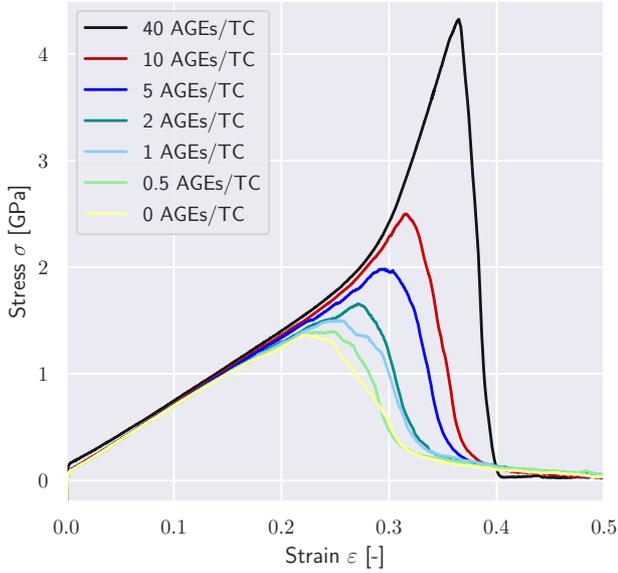

**Figure 2:** Stress-strain curves of tensile tests until rupture of a representative collagen fibril with different AGEs cross-link densities. Simulations were performed with 0 enzymatic cross-links per TC molecule.

Next, the tensile tests are performed in the NVT ensemble, where the rigid bodies at the end of the system are moved apart from each other along the longitudinal axis of the fibril with a constant velocity of 0.0001 Å/fs (= 10m/s) with a time step of $\Delta t = 1$ fs, using a steered-molecular dynamics approach. The applied strain rate is, for computational reasons, considerably faster than commonly used values in experiments. We verified that slower values do not lead to qualitative differences in our simulation results. We use the force calculated between the two groups of atoms being moved apart during the tensile tests to calculate the engineering stress, where the area of the cross-section of the fibril is calculated during the construction of the fibril geometry.

## 3. Results

### 3.1. Fibrilar stress-strain response dependent on AGEs cross-link density

In a first series of simulations, we vary the AGEs cross-link content and investigate its effect on the mechanical properties of the collagen fibril. Our results show that the modeled collagen fibrils present a linear deformation behavior at low strains, i.e. $\varepsilon < \varepsilon_0 \approx 0.15$, where $\varepsilon_0$ defines the limit of the linear elastic regime.(see Fig. 2). This behavior is qualitatively and quantitatively independent of the AGEs density.

The effect of the AGEs content starts to appear for strains beyond this limit, where fibrils with lower cross-link densities (e.g., 0-1 AGEs/TC) show softening mechanisms. The fibril stiffness reduces continuously, transitioning into a reduction of stresses, which indicates the presence of localization in terms of reduction in cross-sectional area, and eventually leading to total failure of the fibril without any load-carrying capacity (Fig. 2). The failure process appears to be overall smooth (no sudden stress drops) and occurs at relatively low stress and strain levels. Fibrils with higher AGEs densities present a very different mechanical behavior (e.g., 10-40 AGEs/TC in Fig. 2). At $\varepsilon > \varepsilon_0$, they present a stiffening, which leads quickly to considerably higher stresses for relatively moderate strain levels. Moreover, failure of these fibrils appears more abrupt, where stresses drop fast and without any prior softening that would indicate an impeding fracture (note sharp peak in stress-strain curves for 40 AGEs/TC in Fig. 2).

These differences in the collagen fibril behavior translate into various mechanical properties that are affected by the AGEs density, and which may result in some of the observed impairments at the tissue level. First, we observe that the peak stress $\sigma_{peak}$ increases with a larger AGEs cross-link content $N_{AGE}$ (see Fig. 3a, b, and Appendix 8.2). This observation is consistent across all modeled fibrils and for any enzymatic cross-link content, as will be discussed in Sec 3.2. Second, the work to failure $W_f = \int \sigma d\varepsilon$ is also increasing with increasing $N_{AGE}$ (see Fig. 3c, d, and Appendix 8.2).

The increases in $\sigma_{peak}$ and $W_f$ are directly associated with the observed stiffening of the fibril at $\varepsilon > \varepsilon_0$. To evaluate the contribution of the stiffened regime, we compute the elastic-to-peak stress difference given by $\Delta \sigma = \sigma_{peak} - \sigma_0$, where $\sigma_0 = \sigma(\varepsilon_0)$ is the stress indicating the start of the stiffened regime. We observe that $\Delta \sigma \approx 0$ for fibrils with low $N_{AGE}$ (see Fig. 3e, f) confirming the absence of this stiffening regime, whereas a higher $N_{AGE}$ leads to an increasing $\Delta \sigma$.

### 3.2. Influence of enzymatic cross-linking

In physiological conditions, AGEs cross-links are built in the process of aging or due to increased glycation levels in diabetic patients' tissue. Therefore, enzymatic cross-links will always be present when AGEs occur but their content may vary. Since the influence of these enzymatic cross-links cannot be neglected, we investigate their effect with a series of simulations by varying their content and type, i.e. divalent or trivalent.

Overall, our results show that trivalent enzymatic cross-links have a larger influence than divalent on key mechanical properties of the collagen fibril (compare top with bottom in Fig. 3). We see that the development of increasing peak stress and increasing work to failure is the same for increasing enzymatic cross-link content independent of cross-link type. Specifically, the contribution of ECLs on the peak stress remains relatively constant across various AGEs cross-link densities $N_{AGE}$, while their influence on the work to failure decreases slightly with increasing $N_{AGE}$.

Particularly important is the effect of enzymatic cross-links on the actuation of the stiffened regime. First, we note that enzymatic cross-links alone cannot cause the stiffened regime, as can be seen by $\Delta\sigma \approx 0$ for $N_{AGE} = 0.5$ CLs/TC (see Fig. 3c). However, at low contents of AGEs cross-links, enzymatic cross-links, and in particular trivalent ones, may cause a stiffening of the collagen fibril that would not occur without them (see $N_{AGE} = 1 - 5$ CLs/TC in Fig. 3e). In fibrils with high $N_{AGE}$, we observe that enzymatic cross-links increase further the stiffened regime by increasing the peak strength, but they do not appear to cause any qualitative differences in the behavior of the fibril.

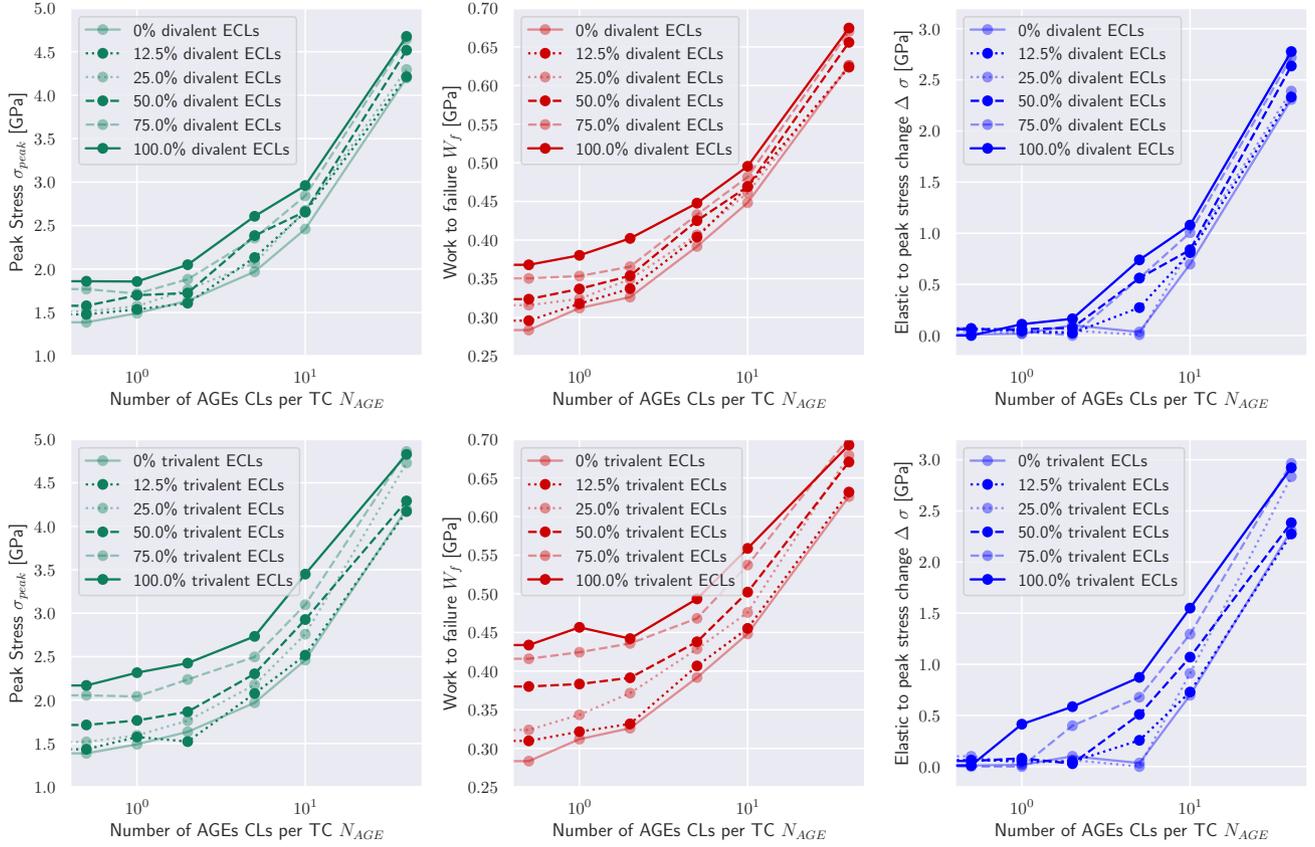

**Figure 3:** Summary of key mechanical properties of collagen fibrils with varying AGEs and enzymatic cross-link content. (a,b) peak stress, (c,d) work to failure and (e,f) $\Delta\sigma$ of collagen fibril with varying AGEs content and (a,c,e) divalent or (b,d,f) trivalent enzymatic cross-links. Results for $N_{AGE} = 0$ not shown but approximately equivalent to $N_{AGE} = 0.5$ CLs/TC. Lines are drawn as guides for the readers' eyes.

### 3.3. Local force distribution in collagen fibril

The observed differences in the global mechanical properties of the collagen fibril are the results of changes in how the force is transmitted through the fibril. Hence, we analyze the force distribution within the TC molecules and the different types of cross-links. Our model shows that the average force within the TC molecules $\bar{F}_{TC}$ follows the stress-strain curves on the global fibril level (compare Fig. 4a with Fig. 2). The same stiffening effect at higher strains for fibrils with $N_{AGE} \geq 5$ CLs/TC is observed, where $\bar{F}_{TC}$ increases after a global strain reaches the critical value of $\varepsilon_0 \approx 0.15$. This shows that the higher loads are mainly carried by the collagen molecules. It also suggests, as can be expected, that the TC molecules are not causing the changes in global behavior but are simply the carrying element in the system.

The average forces in the enzymatic cross-links $\bar{F}_{ECL}$ presents a different behavior. Their maximum values appear to be independent of the cross-link density (see Fig. 4b). This shows that failure of ECLs is not the governing factor of fibrillar failure under the investigated conditions. Furthermore, we note that the ECLs are activated at somewhat higher global strain values for fibrils with higher $N_{AGE}$ (see shift to higher strain values in Fig. 4b). For the linear range $\varepsilon < \varepsilon_0$, where global stress-strain relation is independent of $N_{AGE}$, this shows that the force transmission from on TC molecule to another happens through a different part of the fibril, *i.e.* the AGEs.

The most prominent differences appear in the average forces in AGEs cross-links $\bar{F}_{AGE}$ (see Fig. 4c). Larger $N_{AGE}$ lead to a lower $\bar{F}_{AGE}$ at the same global strain level, with $\bar{F}_{AGE}$ at 40 CLs/TC being less than half of that at 0.5 CLs/TC. These results provide an explanation for the observed stiffening behavior of the collagen fibril. At low $N_{AGE}$, forces transmission between TC molecules within the fibril at $\varepsilon > \varepsilon_0$ happens to some degree through the AGEs and the ECLs but mostly through direct interactions between TC molecules (*e.g.*, friction). Even an increase by a factor 4 from 0.5 CLs/TC to 2 CLs/TC does not cause any significant changes to $\bar{F}_{AGE}$. Consequently, global fibrillar deformation in this strain range and at these $N_{AGE}$ becomes increasingly dominated by intra-fibrillar sliding. This is confirmed by decreasing local strain of the TC molecules at increasing global fibrillar strain (see Fig. 5).

At $N_{AGE} > 2$, however, the lower $\bar{F}_{AGE}$ indicates that the AGEs cross-link become the primary medium for force transmission between TC molecules (same global force is distributed over more cross-links). As a consequence, sliding between TC is mostly oppressed (confirmed by Fig. 5) and global deformation is the direct result of the deformation of the TC molecule. Since these molecules are the stiffest part of the collagen fibril (see Fig. 8), this transfer of deformation from sliding between TC molecules to deformation of the TC molecules explains the observed stiffening of the collagen fibril.

### 3.4. Local origins of collagen fibril failure

Modifications to the deformation mechanisms of collagen fibrils incurred by changes in $N_{AGE}$ are likely causing also alteration to their failure mechanisms. These changes manifest themselves in the timing and quantity of failed AGEs and ECLs cross-links, as well as the breaking of the TC molecules.

First, we note that in the elastic regime $\varepsilon < \varepsilon_0$, the deformation is mostly driven by stretching of the TC molecules (see $\Delta\varepsilon > 0$ Fig. 5b). Consequently, with only very few exceptions, no cross-links are broken in this regime (note almost zero broken AGEs and ECLs even for $\varepsilon = 0.25 > \varepsilon_0$ in Fig. 6&7). As the deformation mechanism transforms into a process dominated by intermolecular sliding, the breaking of cross-links slowly starts to appear. This is expected as sliding strains the cross-links. The failure of cross-links then reduces the resistance against intermolecular sliding, which causes sliding to become increasingly dominent throughout this process (note how $\Delta\sigma$ grows negatively in Fig. 5b).

Our model shows that AGEs are the first cross-links failing in collagen fibrils with relatively low $N_{AGE}$, independent of the ECLs content (see Figs. 6&7). Overall, $40-60\%$ of the AGEs cross-links break until failure of the collagen fibril. The breaking of ECLs occurs at higher strain levels but eventually reach a similar overall failure ratio of 40%. Most importantly, these collagen fibrils with low $N_{AGE}$ do not show any breaking of TC molecules. Hence, it is the ECLs and AGEs cross-links that govern the overall failure of collagen fibrils with low $N_{AGE}$, which confirms our observations from Sec. 3.3 that their failure mechanism is sliding-dominated.

Collagen fibrils with high $N_{AGE}$, however, we observe a different mechanisms. As discussed in Sec. 3.3, the large number of AGEs cross-links act as sliding inhibitors, and fibrillar deformation is dominated by the deformation of the TC molecules. Consequently, less cross-links fail during the deformation of the fibril (see Figs. 6&7). However, the TC molecules break at high fibrilar strain levels (see Figs. 6d&7d). Since the TC molecules are considerably stronger than the cross-links (see Fig. 8) and fail abruptly, the global failure mechanism of the collagen fibril also becomes more abrupt, as we have observed in Fig. 2.

## 4. Discussion
### 4.1. Limitations of numerical model

Our model is a simplification of a collagen fibril, which is a complex biological system. Hence, there are various underlying assumptions that may need further investigation in the future. For instance, we modeled one specific type of AGEs cross-links, namely glucosepane, which is known to be the most abundant one in collagenous tissues (Sell et al., 2005). Other AGEs cross-links have not been taken into account because most of them have not been quantified in tissue so far. Furthermore, the locations of AGEs cross-linking remain largely unknown, which is why we considered a random AGEs distribution along the TC molecule to approximate the amino-acids-based process responsible for cross-link position selection (Gautieri et al., 2014). Apart from AGEs, also enzymatic cross-links display a higher configurational variety than displayed in our study. For example, divalent and trivalent cross-links are not mutually exclusive and co-exist, which was not considered in our model.

The range of AGEs content in our model exceeds considerably their occurrence as observed in measurements with current technology (Sell et al., 2005). While this is true for glucosepane alone, the observed mechanical behavior is likely still to occur for two reasons. First, other non-glucosepane AGEs are also present in collagen fibrils but have been neglected here, and second, the exact mechanical properties of AGEs still remain unknown, and, hence, stiffer and stronger AGEs are likely to cause the stiffening of the fibril at lower AGEs content.

When it comes to collageneous tissues, there is great variability in structure and composition. Collagen type I is not the only collagen responsible for the structural properties. Further, the turn-over, i.e. the physiological rebuilding of tissue is important for the accumulation of AGEs since they are only built during the process of aging or with enhanced exposure increased glycation levels (Verzijl et al., 2000). Depending on the glycation level and exposure, the AGEs density might also vary across the cross-section of the fibril and influence the mechanical properties additionally. Types of AGEs might also vary depending on the tissue type (Eyre et al., 2008). We do not account for these biological variabilities in our model, so the collagen fibril is a generic representative.

Direct and quantitative comparison between our numerical results and experimental measurements on collagen fibrils (Svensson et al., 2018; Eppell et al., 2006; Shen et al., 2008) are restricted by uncertainties in the measurements of fibril geometry caused by experimental procedures. Differences may occur due to the definition of the numerical force field deduced from full-scale simulations (Depalle et al., 2015), which relies on strong assumptions at the atomistic scale and might overestimate the strength of the collagen bonds. However, the absolute value of the force fields is not expected to affect the qualitative behavior of the collagen fibril.

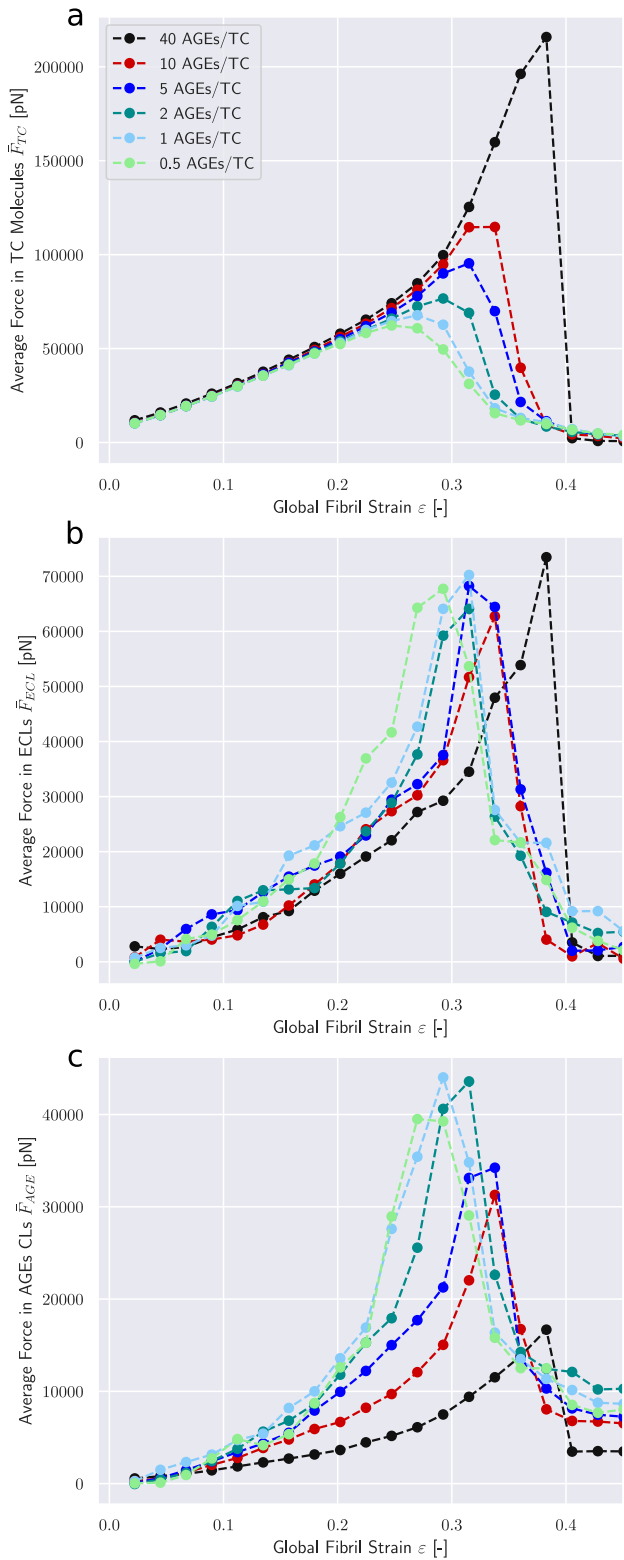

**Figure 4:** Force distribution among bond types within a collagen fibril. Average force in (a) TC molecule bonds, (b) enzymatic cross-links, and (c) AGEs. Fibril with 25 percent ECLs.

Despite all of these limitations, the proposed model captures the relevant mechanical properties of the TC molecule and the various cross-links and provides a valuable qualitative description of the deformation and failure mechanisms inside collagen fibrils. Finally, we note that our model is limited to the scale of collagen fibril and thus we can only analyze the effect of AGEs density on the fibril. Intramolecular AGEs, which are present within the TC molecules or bonded to these, may also reduce the strength of the TC molecules and therefore affect the tissue properties (*e.g.*, strength, stiffness). This AGEs contribution is beyond the scope of the present work and is left for future models.

### 4.2. Physiological implications of results

Collagen fibrils are a main building component of numerous tissues and their behavior directly affects the macroscopic mechanical properties of the tissue. For example in bone, researchers found that an increased glycation level with increased AGEs density is associated with increased brittleness and increased likelihood of fracture (Rubin et al., 2016; Hunt et al., 2019; Vashishth et al., 2001; Tang et al., 2007; Acevedo et al., 2018a, 2015). It was commonly argued that the increased amount of AGEs causes brittleness of bone by inducing a stiffening of the collagen fibril and by preventing fibrillar sliding, which is assumed to be the natural energy dissipation mechanism (Zimmermann et al., 2011; Acevedo et al., 2018b,a). Our results support this perspective of the cause for bone brittleness. In fact, our model provides a clear causal link between increased cross-linking within the collagen fibril and the occurrence of stiffening and strengthening of the fibril. The simulations demonstrate that both of these modifications of the fibrillar deformation mechanisms are caused by inhibited fibrillar sliding between TC molecules due to increased AGEs content. This suggests that energy dissipation is reduced because only little friction occurs, and, consequently the tissue appears to be more brittle. Hence, our results demonstrate that increased fibrillar cross-linking with AGEs is directly responsible for bone brittleness. It is important to note that this brittleness occurs despite an increase in the work to failure, which is not an appropriate descriptor for fracture toughness. In contrast, our results clearly show that fibrils with higher AGEs content fail in a considerably more abrupt manner (see the post-peak slope in Fig. 2).

Aside from the more brittle behavior of the collagen fibril, there are additional mechanisms contributing to an increased brittleness of bone. For instance, collagen might not be the origin of failure in the bone with high AGEs cross-linking due to its increased strength. Instead, fracture initiation might occur in the mineral component that surrounds the collagen. As minerals are known to be brittle, this would inherently increase the overall brittleness of the bone. To evaluate these contributions to impaired material properties of bone, future models should target a scale that combines collagen fibrils and minerals.

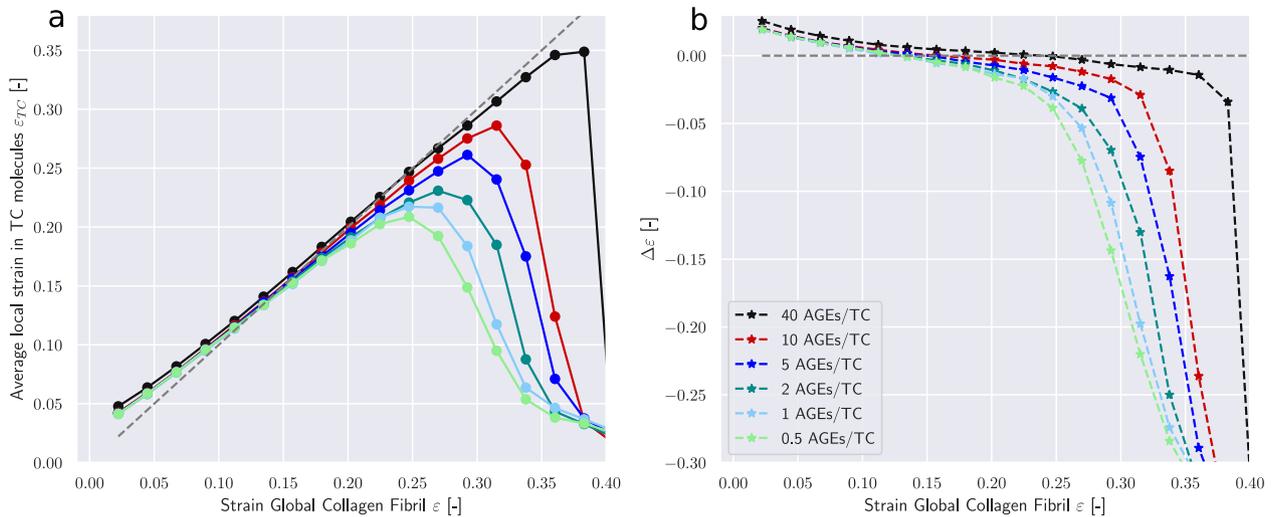

**Figure 5:** Deformation mechanisms in collagen fibril with an ECL content of 75 percent. (a) Local strains in TC molecules $\varepsilon_{TC}$ at global fibrillar strain $\varepsilon$. (b) The difference between global strain an average strain in TC bonds $\Delta\varepsilon = \varepsilon - \varepsilon_{TC}$ as function of global fibrillar strain.

Finally, it is important to note that since these hypotheses for the origin of brittleness in bone with high AGEs content are based purely on numerical simulations, it would be important to provide experimental confirmation. For this, it would be advisable to mechanically test individual collagen fibrils with different cross-link densities. Such experiments would also provide data to better calibrate the numerical model and address some of the limitations mentioned above.

## 5. Conclusion

We performed in-silico tensile tests on collagen fibrils using a coarse-grained molecular simulations of a representative part of the fibril. We analyzed the effect of increased amounts of AGEs non-enzymatic cross-links on the mechanical properties of the collagen fibril. We found that higher AGEs cross-link densities cause increased strength and work to failure of the fibril. Furthermore, we discovered that the collagen fibril presents an increased stiffness for large strains at AGEs cross-link densities that exceed some critical level. We showed that this is valid for fibrils with varying contents of enzymatic cross-links. We demonstrated that the stiffening is the result of the AGEs acting as the main medium for force transfer between the TC molecules, which inhibits intrafibrillar sliding and causes the macroscopic deformation being dominated by the deformation of the (stiffer) TC molecules. Consequently, energy dissipation, which in fibrils with low cross-linking occurs mostly through friction as TC molecules slide against each other, is considerably reduced in collagen fibrils with dense AGEs cross-linking. Hence, our model results provide direct and causal evidence for the link between high AGEs cross-link content and impaired mechanical material properties (*e.g.*, increased brittleness) in collageneous tissue, as commonly observed in aged and diabetic bone.

## 6. Acknowledgement

We are grateful to James L. Rosenberg (Utah), Ihsan Elnunu (Utah) and Dr. Hajar Razi (ETH) for helpful discussions.

## 7. Funding

Research reported in this publication was supported by NIAMS of the National Institutes of Health under award number 1R21AR077881.

## CRediT authorship contribution statement

**Julia Kamml:** Formal analysis, Investigation, Visualization, Writing - Original Draft . **Chun-Yu Ke:** Methodology. **Claire Acevedo:** Conceptualization, Funding acquisition, Writing - Review & Editing . **David S. Kammer:** Conceptualization, Methodology, Funding acquisition, Writing - Review & Editing .

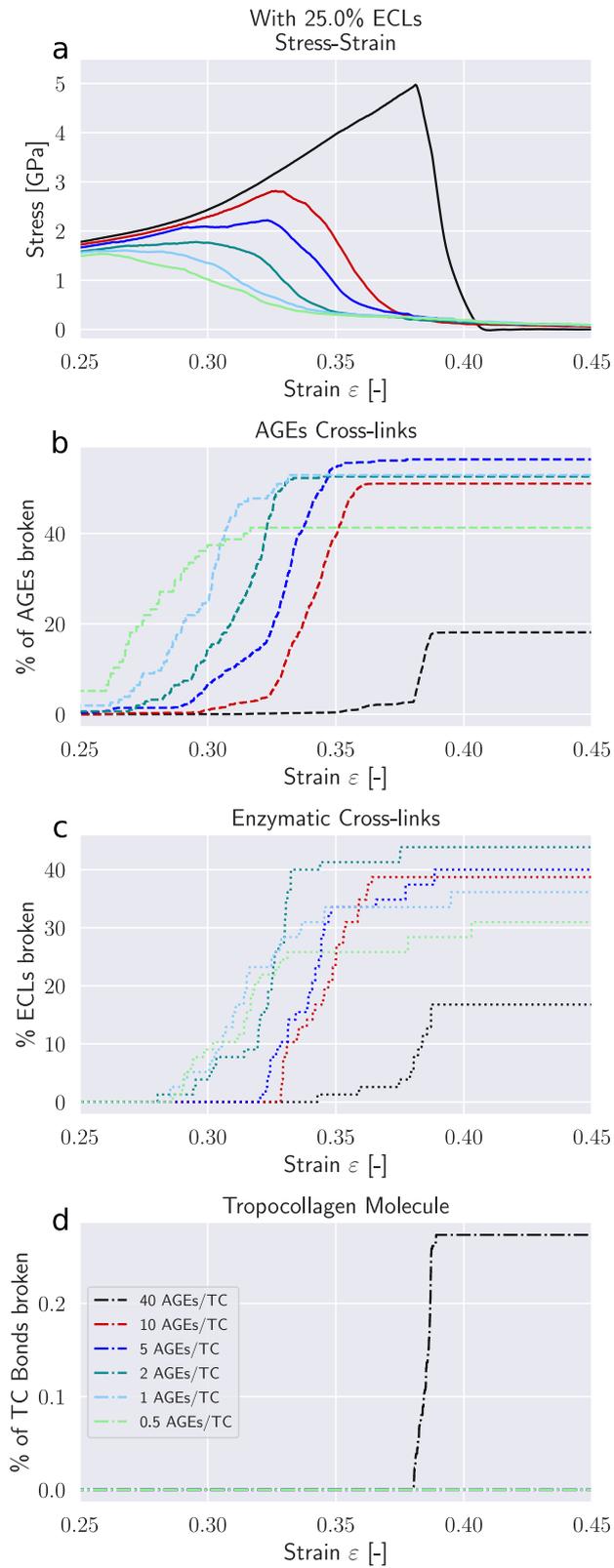
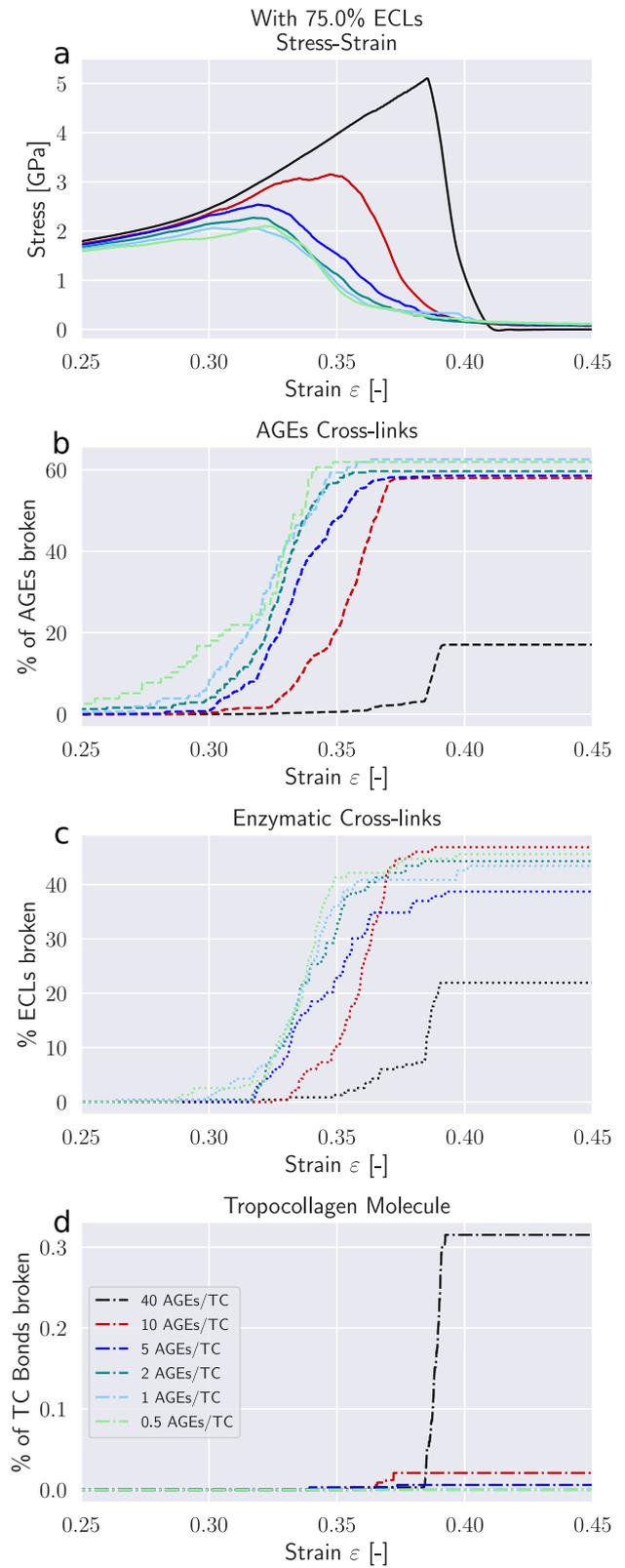

**Figure 6:** Failure of cross-links in collagen fibril with 25 percent ECL. (a) Stress-strain curve of collagen fibril. (b) Percentage of broken AGEs cross-links. (c) Percentage of broken enzymatic cross-links. (d) Percentage of broken bonds in TC molecules.

**Figure 7:** Failure of cross-links in collagen fibril with 75 percent ECL. (a) Stress-strain curve of collagen fibril. (b) Percentage of broken AGEs cross-links. (c) Percentage of broken enzymatic cross-links. (d) Percentage of broken bonds in TC molecules.

The influence of AGEs and enzymatic cross-links on the mechanical properties of collagen fibrils

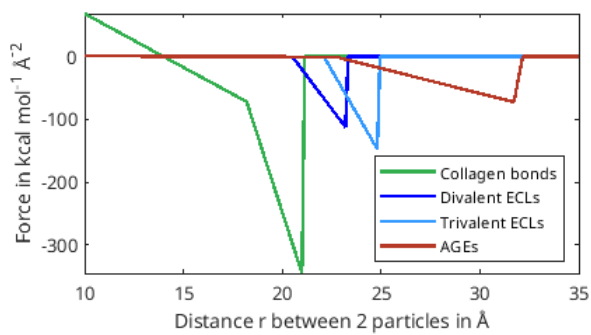

**Figure 8:** Force-distance relation of different bond types

# 8. Appendix
## 8.1. Definition of the bond potentials

The force-distance parameters for collagen bonds in the TC molecules and the enzymatic cross-links are based on Depalle et al. (2015). The mechanical properties of AGEs cross-links were obtained from full-scale molecular dynamics simulations on the geometry of glucosepane using the ReaxxFF protein force-field following Crippa (2013). Figure 8 shows the relation of the three different bond types used in our coarse-grained model, where enzymatic cross-links are either modeled as divalent or trivalent bonds. We used 18.72Å as a general equilibrium distance for inserting cross-links after equilibration, as the bead diameter is defined as 14.72Å and the maximum cross-link length is estimated to be 3.8Å (Gautieri et al., 2014). Cross-links were inserted with a distance criterion in order to stay close to their equilibrium position. Parameters are displayed in Table 1.

## 8.2. Mechanical properties of collagen fibrils in tensile

From Fig. 9, we obtain the mechanical properties from different simulations of destructive tensile tests on collagen fibrils with different AGEs and ECLs densities.

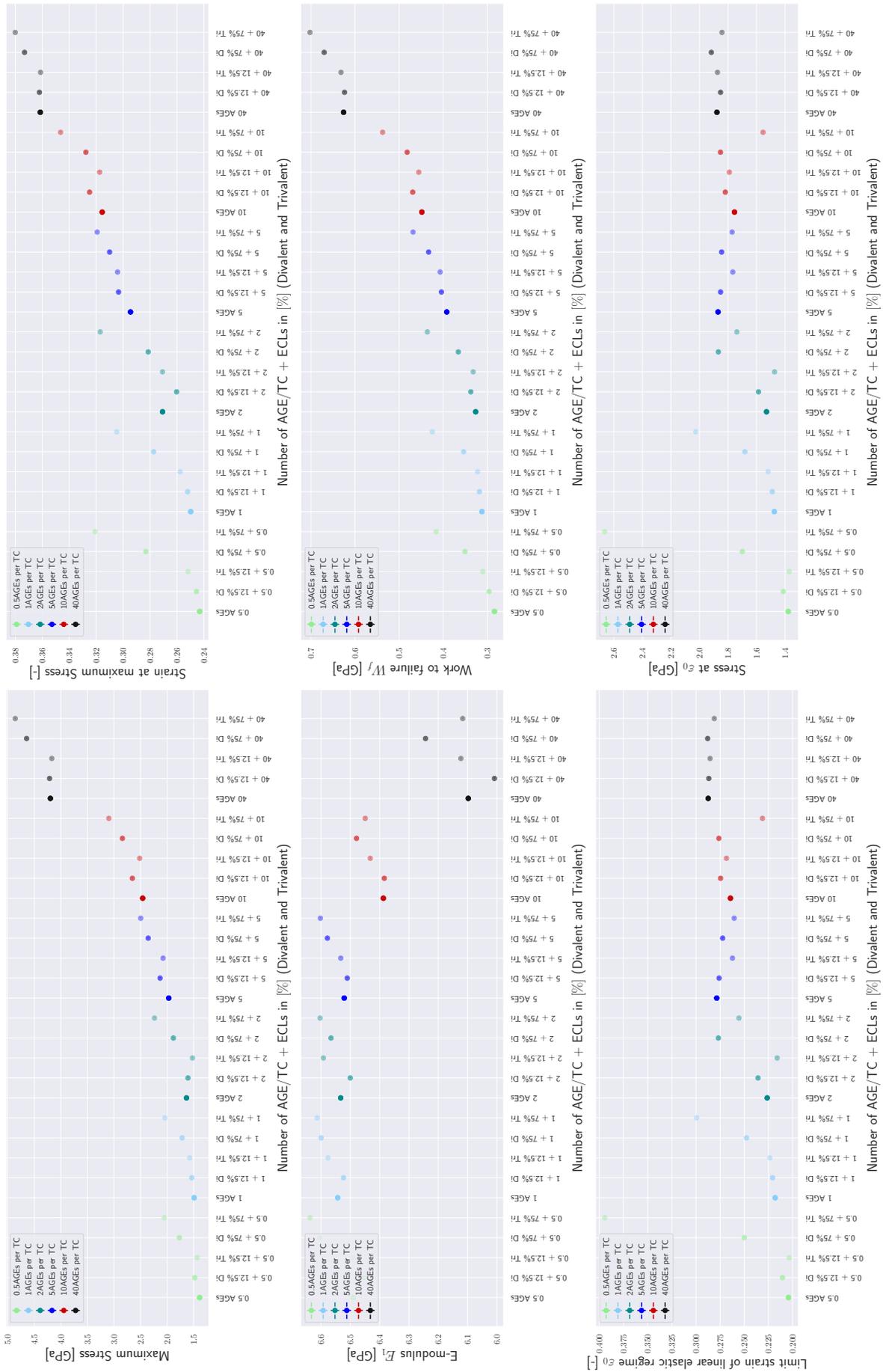

Figure 9: Mechanical properties of collagen fibrils with varying combinations of ECLs and AGEs densities.